\documentclass[11pt,a4paper]{article}
\usepackage{amsmath}
\usepackage[mathcal]{eucal}
\usepackage{latexsym}
\usepackage{amssymb}
\begin{titlepage}
\title{Instanton representation of Plebanski gravity XX. Minisuperspace quantization of gravity coupled to the Klein--Gordon scalar field}
\author{Eyo Eyo Ita III}
\medskip
\input amssym.def
\input amssym.tex

\def \in{\indent}

\begin{document}
\maketitle
\bigskip
\centerline{Department of Applied Mathematics and Theoretical Physics} 
\smallskip
\centerline{Centre for Mathematical Sciences, University of Cambridge, Wilberforce Road}
\smallskip
\centerline{Cambridge CB3 0WA, United Kingdom}
\smallskip
\centerline{eei20@cam.ac.uk} 

\bigskip

\begin{abstract}
In this paper we quantize the Klein--Gordon scalar field coupled to gravity in the instanton representation for spatially homogeneous variables.  The construction provides a well-defined 
Hilbert space of states for generic self-interaction potential V, with a direct link to the semiclassical limit below the Planck scale.  Additionally, we compute the Hamilton's equations of motion for this model, performing various consistency checks on the quantum theory.  
\end{abstract}
\end{titlepage}

\section{Introduction}

In the semiclassical limit below the Planck scale, the effects of quantum gravity are presumably weak in the perturbative regime.  However, the quantum field theory of matter fields in this regime is directly accessible to accelerator experiments.  Since all matter couples to gravity, then one should in principle be able to accomodate the gravitational interactions at the quantum level, even if the effects may appear weak in this limit.  The aim of this paper is to uncover possible mechanisms by which these effects can be magnified from the Planck scale to everyday energy scales.\par
\indent
For this paper we will often use the terminology $\coprod$ to signify the semiclassical limit below the Planck scale.  All of the properties of a theory should in principle be computable from the quantum states of the system.  In the limit of $\coprod$ where gravity is weak, the wavefunction of the universe must be fixed by some sort of boundary condition.  For this boundary condition we will assume that the laws of physics reduce to those in Minkowski spacetime.  The action for a spatially homogeneous Klein--Gordon scalar field in this limit is given by

\begin{eqnarray}
\label{UNIVERSE}
S=\int{dt}\bigl({1 \over 2}\dot{\phi}^2-V\bigr),
\end{eqnarray}

\noindent
where $V=V(\phi)$ is the self-interaction scalar potential.  The classical equation of motion for the scalar field in $\coprod$ is given by

\begin{eqnarray}
\label{UNIVERSE1}
\ddot{\phi}+V^{\prime}=0,
\end{eqnarray}

\noindent
where $V^{\prime}={{dV} \over {d\phi}}$.  The Hamiltonian is given by 

\begin{eqnarray}
\label{UNIVERSE11}
H={{\pi^2} \over 2}+V,
\end{eqnarray}

\noindent 
where $\pi=\dot{\phi}$ is the conjugate momentum for $\phi$.  Upon quantization one promotes the dynamical variables to quantum operators $\phi\rightarrow\hat{\phi}$ and $\pi\rightarrow\hat{\pi}$ satisfying equal-time commutation relations

\begin{eqnarray}
\label{UNIVERSE2}
\bigl[\hat{\phi}(t),\hat{\pi}(t)\bigr]=-i\hbar.
\end{eqnarray}

\noindent
The quantum states of the system $\bigl\vert{f}\bigr>$ satisfy a Schr\"odinger equation

\begin{eqnarray}
\label{UNIVERSE3}
\hat{H}\bigl\vert{f}\bigr>=0,
\end{eqnarray}

\noindent
which in principle can be solved once the potential $V$ is specified.  Since the Hamiltonian is Hermitian, then the quantum states are orthonormal with respect to some measure such that

\begin{eqnarray}
\label{UNIVERSE4}
\bigl<f\bigl\vert{g}\bigr>=\delta_{fg},
\end{eqnarray}

\noindent
and one has the ability to compute expectation values for the system.  We will assume without loss of generality the following semiclassical form for the wavefunction

\begin{eqnarray}
\label{GIVENBY}
\bigl<\phi\bigl\vert{f}\bigr>=\Phi(\phi)=\hbox{exp}\Bigl[{{{l^3} \over \hbar}\int^{\phi}f(\varphi)\delta\varphi}\Bigr].
\end{eqnarray}

\noindent
In the Schr\"odinger representation, then (\ref{GIVENBY}) is annihilated by the Hamiltonian in the form

\begin{eqnarray}
\label{GIVENBY1}
\Bigl({1 \over 2}\Bigl(-(\hbar{l}^{-3})^2{{\partial^2} \over {\partial\phi^2}}+V\Bigr)\Phi(\phi)=0,
\end{eqnarray}

\noindent
where the self-interaction potential is given by

\begin{eqnarray}
\label{GIVENBY2}
V(\phi)={1 \over 2}\Bigl(f^2+{\hbar \over {l^3}}f^{\prime}\Bigr).
\end{eqnarray}

\noindent
It is more convenient to regard the semiclassical matter momentum $f$ as the fundamental, and the potential $V(\phi)$ as a derived quantity.  Then one may use $V(\phi)$ as given by (\ref{GIVENBY2}) as the self-interaction potential for the scalar field in $\coprod$, and one has a closed form solution for (\ref{UNIVERSE3}).\par
\indent
Having determined the form of the solution in $\coprod$, we will now couple the scalar field $\phi$ to gravity and quantize the coupled system.  The link from the coupled theory to $\coprod$ is via the self-interaction 
potential $V(\phi)$, which we assume to be the same in $\coprod$ as it is at the Planck scale.  This potential (\ref{GIVENBY2}) carries the imprints of the physics of $\coprod$ encoded in the label $f$ of the eigenstates in $\coprod$.  By coupling this system to gravity we will extrapolate this physics to the Planck scale.\footnote{Conversely, starting from the physics of fully coupled gravity at the Planck scale, one should in principle be able to compute the limit of this physics in $\coprod$.  In this sense the physics in one regime can serve to constrain physics in the opposite regime.}  In this sense we may be assured that the coupled quantum gravitational theory should produce the correct semiclassical limit since we have imposed this limit as a boundary condition.\par
\indent
In this paper we construct a Hilbert space for the Klein--Gordon scalar field coupled to gravity, using the instanton representation of Plebanski gravity.  We will consider the simplest case of a Klein--Gordon field in minisuperspace to illustrate the algorithm for construction of the states, and also to demonstrate the link from the quantum gravitationally coupled theory to $\coprod$.  Our definition of minisuperspace is unlike the conventional definitions in that it has nothing to do with Bianchi groups.  According to our definition of minisuperspace, we analyse the theory on configurations of spatially constant variables by setting all spatial gradients automatically to zero.  Our starting point for the gravitational part of the model will be the Ashtekar variables, which we will transform into the instanton representation of Plebanski gravity using the CDJ Ansatz $\widetilde{\sigma}^i_a=\Psi_{ae}B^i_e$.\par
\indent
The organization of this paper is as follows.  Section 2 performs the minisuperspace reduction in conjunction with transforming the Ashtekar formalism into the instanton representation.  Here, we set up the canonical structure for the coupled theory.  Section 3 quantizes the theory for vanishing cosmological term.  This is defined as $\Lambda(\phi)=\Lambda+GV(\phi)=0$, where $\Lambda$ is the cosmological constant and $V(\phi)$ is the self-interaction scalar potential.  In this section we establish the gravitational part of the Hilbert space using coherent states.  Section 4 quantizes the theory for nonvanishing cosmological term, constructing an infinite tower of normalizable states for generic self-interaction potential $V$ by expansion about the states of section 3.  The requirement of convergence of the expansion places a constraints on the allowable states, and in this sense they are regarded as restricted states.  In section 6 we perform the expansion for the coupled theory in relation to the inverse of the cosmological term, an expansion which converges without restrictions on the states.  We impose the boundary condition that the resulting states reduce to (\ref{GIVENBY}) in the limit of $\coprod$.  Section 7 computes the Hamilton equations of motion for a prelimiary check of consistency of the classical dynamics against the states determined in the quantum theory.  For notational purposes, lowercase Latin symbols from the beginning of the alphabet $a,b,c,\dots$ denote internal $SU(2)_{-}$ indices, while from the middle of the alphabet $i,j,k,\dots$ denote spatial indices in 3-space $\Sigma$.

\newpage

\section{Reduction to minisuperspace}

\noindent
The 3+1 decomposition of the action for general relativity coupled to a Klein--Gordon scalar field in Ashtekar variables $\phi$ is given by \cite{ASHMATTER}

\begin{eqnarray}
\label{CANO}
S=\int{dt}\int_{\Sigma}d^3x{1 \over G}\widetilde{\sigma}^i_a\dot{A}^a_i+\pi\dot{\phi}
-{N}(\hbox{det}\widetilde{\sigma})^{-1/2}\bigl({1 \over G}H_{grav}+H_{KG}\bigr)-N^iH_i-A^a_0G_a,
\end{eqnarray}

\noindent
where $(A^a_i,\widetilde{\sigma}^i_a)$ are the self-dual Ashtekar $SU(2)_{-}$ connection and the densitized triad \cite{ASH3}, and $(\phi,\pi)$ are the scalar field and its conjugate momentum.  $(N,N^i,A^a_0)$ are the lapse function, 
shift vector and $SU(2)_{-}$ rotation angle and the corresponding constraints include matter contributions.  The gravitational contribution to the Hamiltonian constraint $H_{grav}$ is given by

\begin{eqnarray}
\label{CANOI}
H_{grav}={{\Lambda(\phi)} \over 3}\epsilon_{ijk}\epsilon^{abc}\widetilde{\sigma}^i_a\widetilde{\sigma}^j_b\widetilde{\sigma}^k_c
+\epsilon_{ijk}\epsilon^{abc}\widetilde{\sigma}^i_a\widetilde{\sigma}^j_bB^k_c
\end{eqnarray}

\noindent
where $\Lambda(\phi)=\Lambda+GV(\phi)$, which is a field-dependent cosmological term due to the self interaction potential $V(\phi)$ for the scalar field $\phi$.\footnote{Note that this includes any contribution from the bare cosmological constant $\Lambda$.}  The Ashtekar magnetic field $B^i_a$ is related to 
the connection $A^a_i$ (for nondegenerate $A^a_i$) by

\begin{eqnarray}
\label{EIGEN2}
B^i_a=\epsilon^{ijk}\partial_jA^a_k+{1 \over 2}\epsilon^{ijk}f_{abc}A^b_jA^c_k=\epsilon^{ijk}\partial_jA^a_k+(\hbox{det}A)(A^{-1})^a_i,
\end{eqnarray}

\noindent
where we have used the fact that the structure constants $f_{abc}=\epsilon_{abc}$ for $SU(2)_{-}$, the gauge group for the self-dual Ashtekar variables, are numerically the same as the three dimensional epsilon symbol.  The matter contribution to the Hamiltonian constraint is given 
by \cite{ASHMATTER}

\begin{eqnarray}
\label{CANOI1}
H_{KG}={{\pi^2} \over 2}+{1 \over 2}\widetilde{\sigma}^i_a\widetilde{\sigma}^j_a\partial_i\phi\partial_j\phi.
\end{eqnarray}

\noindent
The diffeomorphism constraint is given by

\begin{eqnarray}
\label{CONOOO}
H_i={1 \over G}\epsilon_{ijk}\widetilde{\sigma}^j_aB^k_a+\pi\partial_i\phi,
\end{eqnarray}

\noindent
which in the full theory possesses a matter contribution containing spatial gradients.  The Gauss' law constraint is given by

\begin{eqnarray}
\label{GOOOSE}
D_i\widetilde{\sigma}^i_a=\partial_i\widetilde{\sigma}^i_a+f_{abc}A^b_j\widetilde{\sigma}^j_c=0,
\end{eqnarray}

\noindent
which does not have a matter contribution since the Klein--Gordon field $\phi$ is a scalar under $SU(2)_{-}$.  We will now perform a reduction to minisuperspace\footnote{Our definition of minisuperspace is that all dynamical variables are spatially constant, depending only on time.  This is obtained by setting all spatial gradients to zero.  We do not use \cite{KODAMA1}, which entails the introduction of Bianchi groups in the 
definition of minisuperspace.} in conjunction with the transformation into the instanton representation of general relativity using the CDJ Ansatz\footnote{The CDJ Ansatz is named after Riccardo Capovilla, John Dell and Ted Jacobson and uses the CDJ matrix $\Psi_{ae}\in{SO}(3,C)\otimes{SO}(3,C)$ to solve the Hamiltonian and diffeomorphism constraints, which are algebraic constraints in the Ashtekar variables \cite{CAP}.}

\begin{eqnarray}
\label{CANO1}
\widetilde{\sigma}^i_a=\Psi_{ae}B^i_e.
\end{eqnarray}
 
\noindent
For minisuperspace as defined, we set the spatial gradients in (\ref{EIGEN2}) to zero obtaining\footnote{In this paper we are considering configurations only where $A^a_i$ is nondegenerate as a 3 
by 3 matrix, hence $\hbox{det}A\neq{0}$.} 

\begin{eqnarray}
\label{EIGEN3}
B^i_a=(\hbox{det}A)(A^{-1})^a_i;~~\hbox{det}B=(\hbox{det}A)^2.
\end{eqnarray}

\noindent
In minisuperspace $\partial_i\phi=0$ and the diffeomorphism constraint (\ref{CONOOO}) becomes

\begin{eqnarray}
\label{FFEE}
\epsilon_{ijk}B^j_aB^k_e\Psi_{ae}=0,
\end{eqnarray}

\noindent
which implies that $\Psi_{ae}=\Psi_{(ae)}$ is symmetric in $a$ and $e$.  In minisuperspace we have $\partial_i\widetilde{\sigma}^i_a=0$ and the relations $A^b_jB^j_e=A^b_j(A^{-1})^j_e(\hbox{det}A)=\delta_{be}(\hbox{det}A)$, whence the Gauss' law constraint (\ref{GOOOSE}) reduces to

\begin{eqnarray}
\label{FFEE1}
f_{abc}A^b_jB^j_e\Psi_{ce}=f_{aec}\Psi_{ce}(\hbox{det}A)=0.
\end{eqnarray}

\noindent
Since $(\hbox{det}A)\neq{0}$ then (\ref{FFEE1}) implies that the antisymmetric part vanishes, or that $\Psi_{ae}$ is symmetric in $a$, $e$.  We already knew this from the diffeomorphism constraint (\ref{FFEE}).  Therefore in minisuperspace, the Gauss' law and the diffeomorphism constraints are redundant and their implementation causes a reduction in $\Psi_{ae}$ by only three instead of six degrees of freedom.  

\par
\indent

\newpage

\section{Establishment of the canonical structure}

\noindent
We have shown that the kinematic constraints in minisuperspace are redundant, and moreover are automatically satisfied when $\Psi_{ae}=\Psi_{(ae)}$ is symmetric in $a$ and $e$.  We therefore eliminate $H_i$ and $G_a$ from the starting action (\ref{CANO}), under the condition that $\Psi_{ae}=\Psi_{(ae)}$, and we have the action on the reduced momentum space of 

\begin{eqnarray}
\label{CANO2}
S=l^3\int{dt}{1 \over G}\Psi_{ae}B^i_e\dot{A}^a_i+\pi\dot{\phi}\nonumber\\
-N(\hbox{det}B)^{-1/2}(\hbox{det}\Psi)^{-1/2}\Bigl[{1 \over G}(\hbox{det}B)\bigl({1 \over 2}Var\Psi+\Lambda(\phi)\hbox{det}\Psi\bigr)+H_{KG}\Bigr],
\end{eqnarray}

\noindent
where we have defined $Var\Psi=(\hbox{tr}\Psi)^2-\hbox{tr}\Psi^2$ and have used (\ref{CANO1}) in (\ref{CANOI}).  Also we have defined by $l$ the characteristic length scale of the universe due to integration of the spatially homogeneous variables over 3-space $\Sigma$.\par
\indent
We will first treat the case of vanishing cosmological term.  For $\Lambda(\phi)=0$, the Hamiltonian constraint is given by\footnote{By $\Lambda(\phi)=0$ we mean that all contributions to the cosmological term are vanishing.  Hence this means also that $V(\phi)=0$, and the scalar field interacts with gravity in the absence of a self-interaction potential.}

\begin{eqnarray}
\label{COOL}
{1 \over 2}(\hbox{det}B)Var\Psi=-GH_{KG},
\end{eqnarray}

\noindent
where $H_{KG}$ is the contribution due to the Klein--Gordon field.  Note that the gravitational contribution depends only on the invariants of $\Psi_{(ae)}$, which can be written entirely in terms of the 
eigenvalues $\lambda_g=(\lambda_1,\lambda_2,\lambda_3)$.  When diagonalizable, the CDJ matrix can be written as a polar decomposition

\begin{displaymath}
\Psi_{ae}=O_{af}(\vec{\theta})
\left(\begin{array}{ccc}
\lambda_1 & 0 & 0\\
0 & \lambda_2 & 0\\
0 & 0 & \lambda_3\\
\end{array}\right)_{fg}
O^{-1}_{ge}(\vec{\theta}),
\end{displaymath}

\noindent
where $O_{af}$ is a complex orthogonal matrix parametrized by three complex angles $\vec{\theta}=(\theta^1,\theta^2,\theta^3)$.  Note that the Hamiltonian constraint (\ref{COOL}) depends only on the invariants, which means that the angles $\vec{\theta}$ do not explicitly appear.  Hence the angles $\vec{\theta}$ are unphysical, which means that only three degrees of freedom in $\Psi_{ae}$ are physically relevant for the theory.  Without loss of generality we thus choose $\Psi_{ae}$ to be a diagonal matrix of gravitational momentum variables, and identify $\Psi_{ae}=\delta_{ae}\lambda_e$ with the eigenvalues.  The variance is then given by

\begin{eqnarray}
\label{EIGEN}
Var\Psi=2\bigl(\Psi_{11}\Psi_{22}+\Psi_{22}\Psi_{33}+\Psi_{33}\Psi_{11}\bigr),
\end{eqnarray}

\noindent
which uses the cyclic property of the trace.\par
\indent
Having made the identification $\Psi_{ae}=Diag(\lambda_1,\lambda_2,\lambda_3)$, for the configuration space we choose a diagonal connection $A^a_i=\delta^a_iA^a_a$ with corresponding magnetic field components

\begin{eqnarray}
\label{MAGNETT}
B^1_1=A^2_2A^3_3;~~B^2_2=A^3_3A^1_1;~~B^3_3=A^1_1A^2_2.
\end{eqnarray}

\noindent
The gravitational contribution to the canonical one form is given by 

\begin{eqnarray}
\label{COONO}
\boldsymbol{\theta}_{grav}={{l^3} \over G}\Psi_{ae}B^i_e\delta{A}^a_i={{l^3} \over G}\Psi_{aa}B^i_a\delta{A}^a_i\nonumber\\
={{l^3} \over G}\bigl(\Psi_{11}B^1_1\delta{A}^1_1+\Psi_{22}B^2_2\delta{A}^2_2+\Psi_{33}B^3_3\delta{A}^3_3\bigr),
\end{eqnarray}

\noindent
which prompts the following definition of variables to obtain globally holonomic coordinates on the instanton representation configuration space $\Gamma_{Inst}$.  First we define densitized momentum space variables

\begin{eqnarray}
\label{JACOBI1}
\widetilde{\Psi}_{11}=\Psi_{11}(A^1_1A^2_2A^3_3);~~
\widetilde{\Psi}_{22}=\Psi_{22}(A^1_1A^2_2A^3_3);~~
\widetilde{\Psi}_{33}=\Psi_{33}(A^1_1A^2_2A^3_3).
\end{eqnarray}

\noindent
In the densitized variables, then (\ref{COONO}) is given by 

\begin{eqnarray}
\label{JACOBI2}
\boldsymbol{\theta}_{grav}={{l^3} \over G}
\biggl(\widetilde{\Psi}_{11}\Bigl({{\delta{A}^1_1} \over {A^1_1}}\Bigr)+
\widetilde{\Psi}_{22}\Bigl({{\delta{A}^2_2} \over {A^2_2}}\Bigr)+
\widetilde{\Psi}_{33}\Bigl({{\delta{A}^3_3} \over {A^3_3}}\Bigr)\biggr).
\end{eqnarray}

\noindent
Next, rewrite (\ref{JACOBI2}) in the form

\begin{eqnarray}
\label{JACOBI3}
\boldsymbol{\theta}_{grav}={{l^3} \over G}
\biggl((\widetilde{\Psi}_{11}-\widetilde{\Psi}_{33}){{\dot{A}^1_1} \over {A^1_1}}+
(\widetilde{\Psi}_{22}-\widetilde{\Psi}_{33}){{\dot{A}^2_2} \over {A^2_2}}\nonumber\\
+\widetilde{\Psi}_{33}\Bigl({{\dot{A}^1_1} \over {A^1_1}}+{{\dot{A}^2_2} \over {A^2_2}}+{{\dot{A}^3_3} \over {A^3_3}}\Bigr)\biggr)
\end{eqnarray}

\noindent
and make the following definitions

\begin{eqnarray}
\label{JACOBI4}
\widetilde{\Psi}_{11}-\widetilde{\Psi}_{33}=\Pi_1;~~\widetilde{\Psi}_{22}-\widetilde{\Psi}_{33}=\Pi_2;~~\widetilde{\Psi}_{33}=\Pi
\end{eqnarray}

\noindent
for the momentum space variables, and

\begin{eqnarray}
\label{JACOBI5}
{{\delta{A}^1_1} \over {A^1_1}}=\delta{X};~~{{\delta{A}^2_2} \over {A^2_2}}=\delta{Y};~~{{\delta{A}^1_1} \over {A^1_1}}+{{\delta{A}^2_2} \over {A^2_2}}+{{\delta{A}^3_3} \over {A^3_3}}=\delta{T}.
\end{eqnarray}

\noindent
for the configuration space variables.  Equation (\ref{JACOBI5}) provides holonomic coordinates $(X,Y,T)\in\Gamma_{Kin}$, given by

\begin{eqnarray}
\label{JACOBI6}
X=\hbox{ln}\Bigl({{A^1_1} \over {a_0}}\Bigr);~~Y=\hbox{ln}\Bigl({{A^2_2} \over {a_0}}\Bigr);~~T=\hbox{ln}\Bigl({{A^1_1A^2_2A^3_3} \over {a_0^3}}\Bigr),
\end{eqnarray}

\noindent
where $a_0$ is a numerical constant of mass dimension $[a_0]=1$.\par
\indent
At the kinematical level,\footnote{The kinematic phase space $\Omega_{Kin}$is defined as the phase space at the level of implemention of the Gauss' law and diffeomorphism constraints and prior to the implementation of the Hamiltonian constraint.} the total symplectic two form is 

\begin{eqnarray}
\label{COONO4}
\boldsymbol{\Omega}_{Kin}={{l^3} \over {\hbar{G}}}\Bigl({\delta\Pi}\wedge{\delta{T}}+{\delta\Pi_1}\wedge{\delta{X}}+{\delta\Pi_2}\wedge{\delta{Y}}\Bigr)
+i{{l^3} \over \hbar}{\delta\pi}\wedge{\delta\phi},
\end{eqnarray}

\noindent
and the mass dimensions of the dynamical variables are given by

\begin{eqnarray}
\label{JACOBI8}
[\Pi_1]=[\Pi_2]=[\Pi]=1;~~[X]=[Y]=[T]=0;~~[\phi]=1;~~[\pi]=2.
\end{eqnarray}

\par
\indent 
Upon substitution of (\ref{JACOBI4}) into (\ref{COOL}) and using (\ref{EIGEN}), the Hamiltonian constraint (\ref{COOL}) reduces to

\begin{eqnarray}
\label{COOLIT}
(\hbox{det}B)(\hbox{det}A)^{-2}\Bigl[3\Pi^2+2(\Pi_1+\Pi_2)\Pi+\Pi_1\Pi_2\Bigr]=-GH_{KG}.
\end{eqnarray}

\noindent
In minisuperspace the prefactors in (\ref{COOLIT}) cancel on account of (\ref{EIGEN3}), and the Hamiltonian constraint for $\Lambda=0$ reduces to

\begin{eqnarray}
\label{EIGEN4}
3\Pi^2+2(\Pi_1+\Pi_2)\Pi+\Pi_1\Pi_2+GH_{KG}=0.
\end{eqnarray}

\noindent
At the classical level this has an explicit solution

\begin{eqnarray}
\label{COOL6}
\Pi={1 \over 3}\Bigl(-(\Pi_1+\Pi_2)\pm\sqrt{\Pi_1^2-\Pi_1\Pi_2+\Pi_2^2-3GH_{KG}}\Bigr).
\end{eqnarray}

\noindent

\newpage

\section{Quantization for vanishing cosmological term}

\noindent
We are now ready to perform a quantization of the kinematic phase space $\Omega_{Kin}$.  Upon quantization, the variables $\Pi$, $\Pi_1$, $\Pi_2$ and $\pi$ will become 
promoted to operators $\hat{\Pi}$, $\hat{\Pi}_1$, $\hat{\Pi}_2$ and $\hat{\pi}$, and $T$, $X$ and $Y$ to operators $\hat{T}$, $\hat{X}$ and $\hat{Y}$  satisfying the nonvanishing equal time commutation relations

\begin{eqnarray}
\label{QUANTUM}
\bigl[\hat{T},\hat{\Pi}\bigr]=\bigl[\hat{X},\hat{\Pi}_1\bigr]=\bigl[\hat{Y},\hat{\Pi}_2\bigr]=\mu;~~\bigl[\hat{\phi},\hat{\pi}]={{i\hbar} \over {l^3}}=i{\mu \over G},
\end{eqnarray}

\noindent
where $\mu=\hbar{G}l^{-3}$ such that $[\mu]=1$.  The wavefunction is determined from the following resolution of the identity

\begin{eqnarray}
\label{QUANTUM3}
I=\int{d\phi}\bigl\vert\phi\bigr>\bigl<\phi\bigr\vert\otimes\int{d\mu}\bigl\vert{X},Y,T\bigr>\bigl<X,Y,T\bigr\vert,
\end{eqnarray}

\noindent
whence the state diagonal in the configuration variables is given by

\begin{eqnarray}
\label{QUANTUM4}
\boldsymbol{\psi}(X,Y,T,\phi)=\bigl<X,Y,T,\phi\bigl\vert\boldsymbol{\psi}\bigr>.
\end{eqnarray}  

\noindent
$\boldsymbol{\psi}$ is holomorphic in $X$, $Y$ and $T$ but not in $\phi$, which is real-valued.  Since the gravitational variables are complex, we choose a measure Gaussian in $X$ and $Y$ to provide normalizable wavefunctions.  This is given by \footnote{Note, we do not include a measure in $T$, because we will interpret $T$ as a time variable on $\Gamma_{Inst}$ and one does not normalize a wavefunction in time.  For the matter variable $\phi$ we have the choice between a Gaussian and a delta functional measure.  We will choose the latter for the purposes of this paper.}

\begin{eqnarray}
\label{NOMRAL}
D\mu(X,Y,\phi)={d\phi}\wedge{dX}\wedge{d\overline{X}}{dY}\wedge{d\overline{Y}}\hbox{exp}\Bigl[-(\vert{X}\vert^2+\vert{Y}\vert^2)\Bigr].
\end{eqnarray}

\noindent
In the functional Schr\"odinger representation, holomorphic in $X$, $Y$ and $T$, the operators act respectively by multiplication 

\begin{eqnarray}
\label{QUANTUM1}
\hat{T}\boldsymbol{\psi}=T\boldsymbol{\psi};~~
\hat{X}\boldsymbol{\psi}=X\boldsymbol{\psi};~~
\hat{Y}\boldsymbol{\psi}=Y\boldsymbol{\psi};~~
\hat{\phi}\boldsymbol{\psi}=\phi\boldsymbol{\psi}
\end{eqnarray}

\noindent
and by differentiation

\begin{eqnarray}
\label{QUANTUM2}
\hat{\Pi}\boldsymbol{\psi}=\mu{\partial \over {\partial{T}}}\boldsymbol{\psi};~~
\hat{\Pi}_1\boldsymbol{\psi}=\mu{\partial \over {\partial{X}}}\boldsymbol{\psi};~~
\hat{\Pi}_2\boldsymbol{\psi}=\mu{\partial \over {\partial{Y}}}\boldsymbol{\psi};~~
\hat{\pi}\boldsymbol{\psi}=-i{\mu \over G}{\partial \over {\partial{\phi}}}\boldsymbol{\psi}.
\end{eqnarray}

\noindent
The total Hamiltonian constraint for $\Lambda(\phi)=0$ becomes promoted to an operator which must annihilate the wavefunction $\boldsymbol{\psi}$.  The matter contribution is given by

\begin{eqnarray}
\label{MATTERCONTRI}
G\hat{H}_{KG}={{G\hat{\pi}^2} \over 2}\longrightarrow-{G \over 2}\Bigl({{\mu^2} \over {G^2}}\Bigr){{\partial^2} \over {\partial\phi^2}}
=-\Bigl({{\mu^2} \over {2G}}\Bigr){{\partial^2} \over {\partial\phi^2}}.
\end{eqnarray}

\noindent
The total classical Hamiltonian constraint for $\Lambda=0$ becomes promoted to a quantum operator constraint

\begin{eqnarray}
\label{QUANTUM5}
3\mu^2\Bigl[{{\partial^2} \over {\partial{T}^2}}+{2 \over 3}\Bigl({\partial \over {\partial{X}}}+{\partial \over {\partial{Y}}}\Bigr){\partial \over {\partial{T}}}
+{1 \over 3}{{\partial^2} \over {\partial{X}\partial{Y}}}-\Bigl({1 \over {6G}}\Bigr){{\partial^2} \over {\partial\phi^2}}\Bigr]\boldsymbol{\psi}=0,
\end{eqnarray}

\noindent
with solution

\begin{eqnarray}
\label{WITHSOLUTION}
\boldsymbol{\psi}=ce^{\alpha{X}+\beta{Y}+\lambda{T}}e^{i{{l^3p\phi} \over \hbar}},
\end{eqnarray}

\noindent
where $c$ is a normalization factor and $\alpha$, $\beta$, $\lambda$ and $p$ satisfy the dispersion relation

\begin{eqnarray}
\label{WITHSOLUTION1}
\lambda^2+{2 \over 3}(\alpha+\beta)\lambda+{1 \over 3}\alpha\beta+{{Gp^2} \over {6\mu^2}}=0\nonumber\\
\longrightarrow\lambda={1 \over 3}\Bigl(-(\alpha+\beta)\pm\sqrt{\alpha^2-\alpha\beta+\beta^2-{{3Gp^2} \over {2\mu^2}}}\Bigr)\equiv\lambda_{\alpha,\beta,p}.
\end{eqnarray}

\noindent
The quantities $\alpha$, $\beta$ and $\lambda$ are dimensionless and the dispersion relation (\ref{WITHSOLUTION1}) has caused a reduction from four continuous labels to three.  The constituents of (\ref{WITHSOLUTION}) can be identified with Schr\"odinger representation states 

\begin{eqnarray}
\label{QUANTUM7}
\bigl<X\bigl\vert\alpha\bigr>=e^{\mu^{-1}\alpha{X}};~~\bigl<Y\bigl\vert\beta\bigr>=e^{\mu^{-1}\beta{Y}};~~
\bigl<\phi\bigl\vert{p}\bigr>=e^{{{il^3} \over \hbar}p\phi};~~\bigl<T\bigl\vert\lambda\bigr>=e^{\mu^{-1}\lambda{T}}.
\end{eqnarray}

\noindent
The observation that (\ref{WITHSOLUTION}) is holomorphic in $X$, $Y$ and $Z$ prompts the identification of the gravitational part of the Hilbert space with that of simple harmonic oscillators.  Let us make the identifications

\begin{eqnarray}
\label{WITHSOLUTION2}
\mu^{-1}\hat{\Pi}=\hat{a}_3;~~\mu^{-1}\hat{\Pi}_1=\hat{a}_1;~~\mu^{-1}\hat{\Pi}_2=\hat{a}_2,
\end{eqnarray}

\noindent
where $\hat{a}_f$ are annihilation operators satisfying commutation relations

\begin{eqnarray}
\label{WITHSOLUTION3}
[\hat{a}_f,\hat{a}^{\dagger}_g]=\delta_{fg},
\end{eqnarray}

\noindent
with annihilation operators $\hat{a}_g^{\dagger}$.  Then the Bargmann representation \cite{BARGAIN} implies the following identifications with the oscillator operators

\begin{eqnarray}
\label{WITHSOLUTION4}
\hat{a}_1^{\dagger}=\hat{X};~~\hat{a}_2^{\dagger}=Y;~~\hat{a}_3^{\dagger}=T.
\end{eqnarray}

\noindent
The oscillator representation provides a unique ket ground state $\bigl\vert{0},0,0\bigr>=\bigl\vert{0}\bigr>\otimes\bigl\vert{0}\bigr>\otimes\bigl\vert{0}\bigr>$ which is annihilated by all annihilation operatos

\begin{eqnarray}
\label{WITHSOLUTION5}
\hat{a}_f\bigl\vert{0},0,0\bigr>=0.
\end{eqnarray}

\noindent
The gravitational part of (\ref{WITHSOLUTION}) can be constructed from a set of coherent states in the spirit of Perelomov \cite{PERELOMOV}

\begin{eqnarray}
\label{WITHSOLUTION7}
\bigl\vert\alpha\bigr>=e^{\alpha\hat{a}_1^{\dagger}-\alpha^{*}\hat{a}_1}\bigl\vert{0}\bigr>;~~
\bigl\vert\beta\bigr>=e^{\beta\hat{a}_2^{\dagger}-\beta^{*}\hat{a}_2}\bigl\vert{0}\bigr>;~~
\bigl\vert\lambda\bigr>=e^{\lambda\hat{a}_3^{\dagger}}\bigl\vert{0}\bigr>,
\end{eqnarray}

\noindent
which correspond to the displacement of the vacuum state in the manifold $(\alpha,\beta)\in{C}_2$.\footnote{This would correspond to coherent states for the (complexified) Heisenberg group with generators $\{a_f,a^{\dagger}_f,1\}$ satisfying the algebra (\ref{WITHSOLUTION3}).  The quotient manifold is then $C_2=C_1\otimes{C}_1$, two copies of the complex plane.}  We have singled out $a_3$ to play the role of a time variable on configuration space, and therefore will not carry out a normalization with respect to this variable.  Equation (\ref{WITHSOLUTION7}) are eigenstates of their respective annihilation operators

\begin{eqnarray}
\label{WITHSOLUTION8}
\hat{a}_1\bigl\vert\alpha\bigr>=\alpha\bigl\vert\alpha\bigr>;~~
\hat{a}_2\bigl\vert\beta\bigr>=\beta\bigl\vert\beta\bigr>;~~
\hat{a}_3\bigl\vert\lambda\bigr>=\lambda\bigl\vert\lambda\bigr>.
\end{eqnarray}

\noindent
The following direct product state is annihilated by the Hamiltonian constraint operator

\begin{eqnarray}
\label{QUANTUM6}
\bigl\vert\lambda_{\alpha,\beta,p},\alpha,\beta,p\bigr>=\bigl\vert\alpha\bigr>\otimes\bigl\vert\beta\bigr>\otimes\bigl\vert{p}\bigr>\otimes\bigl\vert\lambda_{\alpha,\beta,p}\bigr>\in{Ker}\{\hat{H}\}, 
\end{eqnarray}

\noindent
given in the oscillator representation by

\begin{eqnarray}
\label{WITHSOLUTION9}
\hat{H}=\hat{a}_3\hat{a}_3+{2 \over 3}(\hat{a}_1+\hat{a}_2)\hat{a}_3+{1 \over 3}\hat{a}_1\hat{a}_2+{{G\hat{\pi}^2} \over {6\mu^2}}.
\end{eqnarray}

\noindent
The overlap between two states is given by\footnote{We have omitted the $\lambda$ part of the label since it is fixed by the dispersion relation, and therefore redundant.}

\begin{eqnarray}
\label{OVERLAP}
\bigl<\alpha,\beta,p\bigl\vert\alpha^{\prime},\beta^{\prime},p^{\prime}\bigr>
=e^{-{1 \over 2}\vert\alpha-\alpha^{\prime}\vert^2}e^{-{1 \over 2}\vert\beta-\beta^{\prime}\vert^2}\delta(p-p^{\prime}).
\end{eqnarray}

\noindent
There is always a nontrivial overlap between the gravitational parts of the wavefunction, which is a consequence of Gaussian measure needed for the holomorphic representation.  On the other hand, the matter contribution to the inner product is nonvanishing only when the two momentum eigenvalues $p$ and $p^{\prime}$ are the same.  Hence, the states as defined by (\ref{QUANTUM6}) are orthogonal with respect to the matter contribution, and overcomplete with respect to the gravitational contribution.  For vanishing cosmological term the states are in two to one correspondence with the manifold $C_2\otimes{R}$.

\newpage

\section{Incorporation of a nonzero cosmological term. Case 1: Restricted matter states}

\noindent
We will now construct solutions to the Hamiltonian constraint for $\Lambda(\phi)\neq{0}$.  The quantum Hamiltonian constraint is given by\footnote{The factor of $e^{-\hat{a}_3^{\dagger}}$ to the right would in 
the Schr\"odinger representation correspond to an operator ordering for the gravitational part of the constraint, of coordinates to the right of the momenta.}

\begin{eqnarray}
\label{WEWILL}
\biggl[\Bigl(\hat{a}_3\hat{a}_3+{2 \over 3}(\hat{a}_1+\hat{a}_2)\hat{a}_3+{1 \over 3}\hat{a}_1\hat{a}_2+{{G\hat{\pi}^2} \over {6\mu^2}}\Bigr)\nonumber\\
+{{\mu\Lambda(\phi)} \over {3a_0^3}}\hat{a}_3(\hat{a}_3+\hat{a}_1)(\hat{a}_3+\hat{a}_2)e^{-\hat{a}_3^{\dagger}}\biggr]\boldsymbol{\psi}=0
\end{eqnarray}

\noindent
where $\Lambda(\phi)=\Lambda+GV(\phi)$ is the scalar field-dependent cosmological term.  Let us assume a wavefunction of the form

\begin{eqnarray}
\label{WEWILL1}
\boldsymbol{\psi}=\bigl\vert\alpha\bigr>\otimes\bigl\vert\beta\bigr>\otimes\bigl\vert\lambda\bigr>\otimes{\chi}(\phi)
\end{eqnarray}

\noindent
where $\chi$ is a matter-dependent part which remains to be determined.  We can replace the action of $\hat{a}_1$ and $\hat{a}_2$ on (\ref{WEWILL1}) with their eigenvalues, yielding a Hamiltonian constraint 

\begin{eqnarray}
\label{WEWILL2}
\biggl[\Bigl(\hat{a}_3\hat{a}_3+{2 \over 3}(\alpha+\beta)\hat{a}_3+{1 \over 3}\alpha\beta+{{G\hat{\pi}^2} \over {6\mu^2}}\Bigr)\nonumber\\
+{{\mu\Lambda(\phi)} \over {3a_0^3}}\hat{a}_3(\hat{a}_3+\alpha)(\hat{a}_3+\beta)e^{-\hat{a}_3^{\dagger}}\biggr]\boldsymbol{\psi}=0
\end{eqnarray}

\noindent
whence the remaining operators act on $\chi$.  Make the following definition

\begin{eqnarray}
\label{WEWILL3}
\hat{a}_3\hat{a}_3+{2 \over 3}(\alpha+\beta)\hat{a}_3+{1 \over 3}\alpha\beta=(\hat{a}_3+m_{-})(\hat{a}_3+m_{+}),
\end{eqnarray}

\noindent
where $m_{\pm}$ are the roots of (\ref{WEWILL3}), seen as a quadratic polynomial in $a_3$.\par
\indent
Let us make the following definition of dimensionless variables

\begin{eqnarray}
\label{LETUSMAKE}
\rho=\sqrt{6G}\phi;~~{1 \over {U(\rho)}}={{\mu\Lambda(\rho)} \over {3a_0^3}}.
\end{eqnarray}

\noindent
We will solve the Hamiltonian constraint (\ref{WEWILL2}) by Lippman--Schwinger type expansion about the $\Lambda(\phi)=0$ states

\begin{eqnarray}
\label{WEWILL5}
\bigl\vert\lambda,p\bigr>\in{Ker}\{-\partial_{\rho}^2+(\hat{a}_3+m_{-})(\hat{a}_3+m_{+})\},
\end{eqnarray}

\noindent
where we have suppressed the $(\alpha,\beta)$ part of the label to avoid cluttering up the notation.  First write the Hamiltonian constraint as

\begin{eqnarray}
\label{WEWILL6}
\bigl(-\partial_{\rho}^2+(\hat{a}_3+m_{-})(\hat{a}_3+m_{+})\bigr)\chi=-{1 \over {U(\rho)}}\hat{a}_3(\hat{a}_3+\alpha)(\hat{a}_3+\beta)e^{-\hat{a}_3^{\dagger}}\chi.
\end{eqnarray}

\noindent
Inverting the operator on the left hand side, we have

\begin{eqnarray}
\label{WEWILL7}
\chi=\bigl\vert\lambda,p\bigr>-\bigl(-\partial_{\rho}^2+(\hat{a}_3+m_{-})(\hat{a}_3+m_{+})\bigr)^{-1}
{1 \over {U(\rho)}}\hat{a}_3(\hat{a}_3+\alpha)(\hat{a}_3+\beta)e^{-\hat{a}_3^{\dagger}}\chi
\end{eqnarray}

\noindent
where we have factored out the part of the states dependent on $\alpha$ and $\beta$.  Equation (\ref{WEWILL7}) can be rearranged to read

\begin{eqnarray}
\label{WEWILL8}
(1+\hat{q})\chi=\bigl\vert\lambda,p\bigr>
\end{eqnarray}

\noindent
with solution

\begin{eqnarray}
\label{WEWILL81}
\chi=(1+\hat{q})^{-1}\bigl\vert\lambda,p\bigr>
=\Bigl(1-\hat{q}+\hat{q}^2-\hat{q}^3+\dots\Bigr)\bigl\vert\lambda,p\bigr>
\end{eqnarray}

\noindent
where we have defined

\begin{eqnarray}
\label{WEWILL9}
\hat{q}=\bigl(-\partial_{\rho}^2+(\hat{a}_3+m_{-})(\hat{a}_3+m_{+})\bigr)^{-1}
{1 \over {U(\rho)}}\hat{a}_3(\hat{a}_3+\alpha)(\hat{a}_3+\beta)e^{-\hat{a}_3^{\dagger}}.
\end{eqnarray}

\noindent
Equation (\ref{WEWILL81}) is an infinite operator expansion, but we will see that it has a well-defined action on the states $\bigl\vert\lambda,p\bigr>$.  The action of $\hat{q}$ is given by

\begin{eqnarray}
\label{WEWILL10}
\hat{q}\bigl\vert\lambda,p\bigr>
=\bigl(-\partial_{\rho}^2+(\hat{a}_3+m_{-})(\hat{a}_3+m_{+})\bigr)^{-1}
{1 \over {U(\rho)}}\hat{a}_3(\hat{a}_3+\alpha)(\hat{a}_3+\beta)e^{-\hat{a}_3^{\dagger}}\bigl\vert\lambda,p\bigr>\nonumber\\
=\bigl(-\partial_{\rho}^2+(\hat{a}_3+m_{-})(\hat{a}_3+m_{+})\bigr)^{-1}
{1 \over {U(\rho)}}\hat{a}_3(\hat{a}_3+\alpha)(\hat{a}_3+\beta)\bigl\vert\lambda-1,p\bigr>\nonumber\\
=\Bigl({{(\lambda-1)(\lambda+\alpha-1)(\lambda+\beta-1)} \over {-\partial_{\rho}^2+(\lambda+m_{-}-1)(\lambda+m_{+}-1)}}\Bigr)
{1 \over {U(\rho)}}\bigl\vert\lambda-1,p\bigr>.
\end{eqnarray}

\noindent
Repeating this $n$ times we have

\begin{eqnarray}
\label{RESTRICT5}
\hat{q}^n\bigl\vert\lambda,p\bigr>
=\biggl({{\prod_{k=1}^n(\lambda-k)(\lambda+\alpha-k)(\lambda+\beta-k)} 
\over {\prod_{k=1}^n\bigl(-\partial^2_{\rho}+\bigl((\lambda-k)^2+{2 \over 3}(\alpha+\beta)(\lambda-k)+{1 \over 3}\alpha\beta\bigr)\bigr)U(\rho)}}\biggr)
\bigl\vert\lambda-n,p\bigr>.
\end{eqnarray}

\noindent

\subsection{Sufficient condition for convergence}

It is not hard to see that (\ref{RESTRICT5}) substituted into (\ref{WEWILL81}) will yield an infinite series generically with a radius of convergence of zero.  In order to obtain finite wavefunctions, the series must be required to terminate a finite order.  A sufficient condition for this is that the numerator of (\ref{RESTRICT5}) be zero

\begin{eqnarray}
\label{BEZERO}
(\lambda-N)(\lambda+\alpha-N)(\lambda+\beta-N)=0
\end{eqnarray}

\noindent
for some $N$, where $\lambda$ satisfies the dispersion relation

\begin{eqnarray}
\label{BEZERO1}
\lambda^2+{2 \over 3}(\alpha+\beta)\lambda+{1 \over 3}\alpha\beta+{{Gp^2} \over {6\mu^2}}=0.
\end{eqnarray}

\noindent
The imposition of convergence will have the effect of eliminating one continous label in favor of a discrete label, which requires that a choice be made.  We require that the continuous gravitational labels $(\alpha,\beta)\in{C}_2$ remain intact, which fixes the matter momentum $p$.  This leads to three different values of $p$ for each pair $(\alpha,\beta)$.  For $\lambda-N=0$ we have

\begin{eqnarray}
\label{HAMEEL21}
p=i\mu\sqrt{{6 \over G}}\Bigl(N^2+{2 \over 3}(\alpha+\beta)N+{1 \over 3}\alpha\beta\Bigr)^{1/2}\equiv{p}_3(\alpha,\beta;N).
\end{eqnarray}
 
\noindent
For $\lambda+\alpha-N=0$ we have

\begin{eqnarray}
\label{HAMEEL22}
p=i\mu\sqrt{{6 \over G}}\Bigl((N-\alpha)^2+{2 \over 3}(\alpha+\beta)(N-\alpha)+{1 \over 3}\alpha\beta\Bigr)^{1/2}\equiv{p}_1(\alpha,\beta;N).
\end{eqnarray}

\noindent
and for $\lambda+\beta-N=0$ we have

\begin{eqnarray}
\label{HAMEEL13}
p=i\mu\sqrt{{6 \over G}}\Bigl((N-\beta)^2+{2 \over 3}(\alpha+\beta)(N-\beta)+{1 \over 3}\alpha\beta\Bigr)^{1/2}\equiv{p}_2(\alpha,\beta;N).
\end{eqnarray}

\noindent
The matter momentum $p_I=p_I(\alpha,\beta;N)$ for $I=1,2,3$ has inherited the gravitational labels $(\alpha,\beta)$ and has become quantized according to the discrete label $N$ in order to produce convergent states.  Likewise the label $\lambda=\lambda(\alpha,\beta;N)$ for the time-dependent part of the state (T-dependent) has acquired the same labels.  Note that $p$ can take on three possible values for each $\gamma\equiv(\alpha,\beta;N)$, which corresponds to three possible states per label.  The states are given by

\begin{eqnarray}
\label{THESTATESARE}
\boldsymbol{\psi}=\bigl\vert\alpha,\beta\bigr>\otimes\chi_{\gamma}(T,\phi);~~\gamma\equiv(\alpha,\beta,N)
\end{eqnarray}

\noindent
where $\chi_{\gamma}(T,\phi)$ is the part dependent on the time variable $T$ and the matter field $\phi$.  The result is that there is an infinite tower of states, which are in three to one correspondence with points on the manifold 
$C_2\otimes{Z}$, where $Z$ corresponds to the integers.

\newpage

\section{Case II: Unrestricted states}

We have seen that an attempt to expand the coupled state in powers of the cosmological term has led to an infinite tower of states restricted by the condition of convergence of the expansion.  Let us rewrite the Hamiltonian constraint (\ref{WEWILL}) in the following form\footnote{Note that we have transferred the operator $e^{\hat{a}_3^{\dagger}}$ from the cubic to the quadratic momentum term.}

\begin{eqnarray}
\label{KEEPINGTHIS}
\biggl[{{3a_0^3} \over {\mu\Lambda(\phi)}}\Bigl(\hat{a}_3\hat{a}_3+{2 \over 3}(\hat{a}_1+\hat{a}_2)\hat{a}_3+{1 \over 3}\hat{a}_1\hat{a}_2+{{G\hat{\pi}^2} \over {6\mu^2}}\Bigr)e^{\hat{a}_3^{\dagger}}\nonumber\\
+\hat{a}_3(\hat{a}_3+\hat{a}_1)(\hat{a}_3+\hat{a}_2)\biggr]\boldsymbol{\psi}=0
\end{eqnarray}

\noindent
where $\Lambda(\phi)=\Lambda+GV(\phi)$ is the scalar field-dependent cosmological term.  Let us assume a wavefunction of the form

\begin{eqnarray}
\label{KEEPINGTHIS1}
\boldsymbol{\psi}=\bigl\vert\alpha\bigr>\otimes\bigl\vert\beta\bigr>\otimes\bigl\vert\lambda\bigr>\otimes{\Phi}(\phi)
\end{eqnarray}

\noindent
We can replace the action of $\hat{a}_1$ and $\hat{a}_2$ on (\ref{KEEPINGTHIS1}) with their eigenvalues on the state, yielding a Hamiltonian constraint 

\begin{eqnarray}
\label{KEEPINGTHIS2}
\biggl[{{3a_0^3} \over {\mu\Lambda(\phi)}}\Bigl(\hat{a}_3\hat{a}_3+{2 \over 3}(\alpha+\beta)\hat{a}_3+{1 \over 3}\alpha\beta+{{G\hat{\pi}^2} \over {6\mu^2}}\Bigr)e^{\hat{a}_3^{\dagger}}\nonumber\\
+\hat{a}_3(\hat{a}_3+\alpha)(\hat{a}_3+\beta)\biggr]\boldsymbol{\psi}=0
\end{eqnarray}

\noindent
whence the remaining operators act on $\Phi$.  We will now be performing a Lippman--Schwinger type expansion about reference wavefunctions 

\begin{eqnarray}
\label{KEEPINGTHIS3}
\{\bigl\vert\alpha,\beta,0\bigr>,
\bigl\vert\alpha,\beta,-\alpha\bigr>,
\bigl\vert\alpha,\beta,-\beta\bigr>\}\otimes\Phi(\phi)\in{Ker}\{\hat{a}_3(\hat{a}_3+\alpha)(\hat{a}_3+\beta)\},
\end{eqnarray}

\noindent
These states have the following Schr\"odinger representation

\begin{eqnarray}
\label{DEAL51}
\bigl\vert0,\alpha,\beta\bigr>=e^{\alpha{X}+\beta{Y}}\Phi(\phi)\equiv\boldsymbol{\psi}_0;\nonumber\\
\bigl\vert-\alpha,\alpha,\beta\bigr>=e^{\alpha{(X-T)}+\beta{Y}}\Phi(\phi)\equiv\boldsymbol{\psi}_1;\nonumber\\
\bigl\vert-\beta,\alpha,\beta\bigr>=e^{\alpha{X}+\beta{(Y-T)}}\Phi(\phi)\equiv\boldsymbol{\psi}_2,
\end{eqnarray}

\noindent
where $\Phi(\phi)$ is the matter contribution.  The physical interpretations are as follows: If we view $T$ as a time variable on configuration space $\Gamma_{Inst}$, then $\boldsymbol{\psi}_0$ is a timeless state, 
and $\boldsymbol{\psi}_1$ and $\boldsymbol{\psi}_2$ correspond to the motion of a free particle plane travelling in a two dimensional configuration space.  Note that $\Phi(\phi)$ can be chosen arbitrarily, since 
it is annihilated by $\hat{a}_3$.  But to make contact with sub-Planckian physics, we chose this contribution to have the physical interpretation of (\ref{GIVENBY}), which reproduces the correct wavefunction in the observable limit in $\coprod$.  The dynamics of quantum gravity should extrapolate this sub-Planckian input to all scales including the Planck scale.  Let us now write the Hamiltonian constraint in the form

\begin{eqnarray}
\label{KEEPINGTHIS4}
\hat{a}_3(\hat{a}_3+\alpha)(\hat{a}_3+\beta)\boldsymbol{\psi}
=-U(\rho)\Bigl(-\partial_{\rho}^2+(\hat{a}_3+m_{-})(\hat{a}_3+m_{+})\Bigr)e^{\hat{a}_3^{\dagger}}\boldsymbol{\psi}
\end{eqnarray}

\noindent
where we have made the definition

\begin{eqnarray}
\label{MADETHE}
U(\rho)={{3a_0^3} \over {\mu\Lambda(\phi)}}={{3a_0^3} \over {\mu(\Lambda+V(\phi))}} 
\end{eqnarray}

\noindent
with $V(\phi)$ given by (\ref{GIVENBY2}).  Inverting the operator on the left hand side we obtain

\begin{eqnarray}
\label{KEEPINGTHIS5}
\boldsymbol{\psi}=\bigl\vert\lambda\bigr>\otimes\Phi(\rho)\nonumber\\
-U(\rho)\bigl(\hat{a}_3(\hat{a}_3+\alpha)(\hat{a}_3+\beta)\bigr)^{-1}\Bigl(-\partial_{\rho}^2+(\hat{a}_3+m_{-})(\hat{a}_3+m_{+})\Bigr)e^{\hat{a}_3^{\dagger}}\boldsymbol{\psi}
\end{eqnarray}

\noindent
where we have used the shorthand notation $\bigl\vert\lambda\bigr>\otimes\Phi$ for (\ref{KEEPINGTHIS3}).  This can be re-arranged to read

\begin{eqnarray}
\label{KEEPINGTHIS6}
(1+\hat{q})\boldsymbol{\psi}=\bigl\vert\lambda\bigr>\otimes\Phi(\rho)
\end{eqnarray}

\noindent
with solution

\begin{eqnarray}
\label{KEEPINGTHIS7}
\boldsymbol{\psi}=(1+\hat{q})^{-1}\bigl\vert\lambda\bigr>\otimes\Phi(\rho)
=\Bigl(1-\hat{q}+\hat{q}^2-\hat{q}^3+\dots\Bigr)\bigl\vert\lambda\bigr>\otimes\Phi(\rho)
\end{eqnarray}

\noindent
where we have defined

\begin{eqnarray}
\label{KEEPINGTHIS8}
\hat{q}=\bigl(\hat{a}_3(\hat{a}_3+\alpha)(\hat{a}_3+\beta)\bigr)^{-1}
U(\rho)\bigl(-\partial_{\rho}^2+(\hat{a}_3+m_{-})(\hat{a}_3+m_{+})\bigr)e^{\hat{a}_3^{\dagger}}.
\end{eqnarray}

\noindent
Equation (\ref{KEEPINGTHIS8}) is an infinite operator expansion.  The action of $\hat{q}$ on the reference state is given by

\begin{eqnarray}
\label{KEEPINGTHIS9}
\hat{q}\bigl\vert\lambda\bigr>\otimes\Phi(\rho)\nonumber\\
=\bigl(\hat{a}_3(\hat{a}_3+\alpha)(\hat{a}_3+\beta)\bigr)^{-1}
U(\rho)\bigl(-\partial_{\rho}^2+(\hat{a}_3+m_{-})(\hat{a}_3+m_{+})\bigr)e^{\hat{a}_3^{\dagger}}
\bigl\vert\lambda\bigr>\otimes\Phi(\rho)\nonumber\\
=\bigl(\hat{a}_3(\hat{a}_3+\alpha)(\hat{a}_3+\beta)\bigr)^{-1}
U(\rho)\bigl(-\partial_{\rho}^2+(\hat{a}_3+m_{-})(\hat{a}_3+m_{+})\bigr)
\bigl\vert\lambda+1\bigr>\otimes\Phi(\rho)\nonumber\\
=U(\rho)\Bigl({{-\partial_{\rho}^2+(\lambda+m_{-}+1)(\lambda+m_{+}+1)} \over {(\lambda+1)(\lambda+\alpha+1)(\lambda+\beta+1)}}\Bigr)
\bigl\vert\lambda+1\bigr>\otimes\Phi(\rho).
\end{eqnarray}

\noindent
Repeating this $n$ times we have

\begin{eqnarray}
\label{RESTRICT55}
\hat{q}^n\bigl\vert\lambda\bigr>\otimes\Phi(\rho)\nonumber\\
=\biggl({{\prod_{k=1}^nU(\rho)\bigl(-\partial^2_{\rho}+\bigl((\lambda+k)^2+{2 \over 3}(\alpha+\beta)(\lambda+k)+{1 \over 3}\alpha\beta\bigr)\bigr)} 
\over {\prod_{k=1}^n(\lambda+k)(\lambda+\alpha+k)(\lambda+\beta+k)}}\biggr)
\bigl\vert\lambda+n\bigr>\otimes\Phi(\rho)\nonumber\\
\equiv{Q}^n_{\alpha,\beta,\lambda}(\rho)\bigl\vert\lambda+n\bigr>\otimes\Phi(\rho).
\end{eqnarray}

\noindent
To obtain the full solution on substitutes (\ref{RESTRICT55}) into (\ref{KEEPINGTHIS7}), which yields

\begin{eqnarray}
\label{WHICHYIELDS}
\boldsymbol{\psi}=\sum_{n=0}^{\infty}\equiv{Q}^n_{\alpha,\beta,\lambda}(\rho)\bigl\vert\lambda+n\bigr>\otimes\Phi(\rho)
\end{eqnarray}

\noindent
Since the denominator of each term of (\ref{RESTRICT55}) exceeds the numerator, which for large $n$ makes $Q^n_{\alpha,\beta,\lambda}$ go as ${1 \over {n!}}$, then we should expect the infinite series (\ref{WHICHYIELDS}) to converge for generic $V(\phi)$.  Another item of note, in contrast to the states (\ref{THESTATESARE}), is that $\lambda$ is now an independent degree of freedom from $(\alpha,\beta)$, as is the matter part of the wavefunctions $\Phi$.  Restoring the gravitational labels and putting $\bigl\vert\lambda\bigr>$ in the Schr\"odinger representation, we can write the solution as

\begin{eqnarray}
\label{WHICHYIELDS1}
\boldsymbol{\psi}_{\alpha,\beta}(\phi,T)=\sum_{n=0}^{\infty}\equiv{Q}^n_{\alpha,\beta,\lambda}(\phi)e^{nT}\bigl\vert\lambda,\alpha,\beta\bigr>\otimes\Phi(\rho).
\end{eqnarray}

\noindent
The result is that (\ref{WHICHYIELDS1}) is a solution of the quantum Hamiltonian constraint with a well-defined semiclassical limit and a direct link to $\coprod$, the limit below the Planck scale, for generic self-interaction potential $V(\phi)$.  The gravitational part of the state has the same labels as in (\ref{KEEPINGTHIS3}),\footnote{This is because we have expanded the solution about these states.} and the matter part is not restricted by the gravitational degrees of freedom.

\newpage

\section{Classical dynamics}

Having constructed the quantum states, we would like to verify consistency with the classical dynamics.  We have used a basis of coherent harmonic oscillator states for the gravitational degrees of freedom.  Consistency would dictate that these states preserve their coherent nature under Hamiltonian evolution.  To verify this we will transform the gravitational variables back into the Schr\"odinger representation and analyse the Hamiltonian dynamics.  The Hamiltonian can be written in the following form

\begin{eqnarray}
\label{DYNAMICS}
\boldsymbol{H}={N \over {\sqrt{\hbox{det}\widetilde{\sigma}}}}\Bigl({1 \over G}(\hbox{det}B)\bigl({1 \over 2}Var\Psi+\Lambda\hbox{det}\Psi\bigr)+H_{KG}\Bigr)\nonumber\\
=N\sqrt{\hbox{det}\widetilde{\sigma}}\Bigl({1 \over {(\hbox{det}\Psi)(\hbox{det}B)}}\Bigr)
\Bigl({1 \over G}(\hbox{det}B)\bigl({1 \over 2}Var\Psi+\Lambda\hbox{det}\Psi\bigr)+H_{KG}\Bigr),
\end{eqnarray}

\noindent
where $H_{KG}={{\pi^2} \over 2}$ and where we have used the CDJ Ansatz (\ref{CANO1}).  The next step is to reduce (\ref{DYNAMICS}) to minisuperspace using $\hbox{det}B=(\hbox{det}A)^2$, followed by densitization of the CDJ matrix

\begin{eqnarray}
\label{DYNAMICS1}
\widetilde{\Psi}_{ae}=(\hbox{det}A)\Psi_{ae}.
\end{eqnarray}

\noindent
The result of applying these steps to (\ref{DYNAMICS}) yields

\begin{eqnarray}
\label{DYNAMICS2}
\boldsymbol{H}
=N(\hbox{det}A)\sqrt{\hbox{det}\Psi}\Bigl({1 \over G}\bigl(\Lambda+\hbox{tr}\Psi^{-1}\bigr)+{1 \over {\hbox{det}\Psi}}(\hbox{det}A)^{-2}H_{KG}\Bigr)\nonumber\\
=N(\hbox{det}A)^{-1/2}\sqrt{\hbox{det}\widetilde{\Psi}}
\Bigl({1 \over G}\bigl(\Lambda+(\hbox{det}A)\hbox{tr}\widetilde{\Psi}^{-1}\bigr)+{{(\hbox{det}A)} \over {\hbox{det}\widetilde{\Psi}}}H_{KG}\Bigr).
\end{eqnarray}

\noindent
We will now restrict $\widetilde{\Psi}_{ae}$ to a diagonal matrix given by

\begin{displaymath}
\widetilde{\Psi}_{ae}=
\left(\begin{array}{ccc}
\lambda & 0 & 0\\
0 & \lambda+\alpha & 0\\
0 & 0 & \lambda+\beta\\
\end{array}\right)
.
\end{displaymath}

\noindent
Recall that $\alpha$, $\beta$ and $\lambda$ in the quantum theory were labels for the gravitational part of the state.  For the purposes of these section we will use the same symbols (which here have an additional numerical factor of $\mu$ with $[\mu]=1$), in order to show clearly the link between the degrees of freedom that have been quantized and their corresponding classical dynamics.  Along these same lines, we will use the same configuration space variables as in the quantum theory

\begin{eqnarray}
\label{CONFIGURATION}
A^1_1=a_0e^X;~~A^2_2=a_0e^Y;~~(\hbox{det}A)=A^1_1A^2_2A^3_3=a_0^3e^T.
\end{eqnarray}

\par
\indent
Next, we will bring out a factor of $(\hbox{det}A)$ from the large brackets in (\ref{DYNAMICS2}), 

\begin{eqnarray}
\label{DYNAMICS3}
\boldsymbol{H}
={1 \over G}Na_0^{3/2}e^{T/2}\sqrt{\lambda(\lambda+\alpha)(\lambda+\beta)}\nonumber\\
\biggl[\Bigl(\Bigl({\Lambda \over {a_0^3}}\Bigr)e^{-T}
+{1 \over \lambda}+{1 \over {\lambda+\alpha}}+{1 \over {\lambda+\beta}}\Bigr)
+{1 \over {\lambda(\lambda+\alpha)(\lambda+\beta)}}\Bigl({{G\pi^2} \over 2}\Bigr)\biggr],
\end{eqnarray}

\noindent
where we have used $(\hbox{det}A)=a_0^3e^T$ from (\ref{CONFIGURATION}).  The Hamilton's equation of motion for the gravitational configuration variables are given by

\begin{eqnarray}
\label{DYNAMICS4}
\dot{X}=G{{\partial\boldsymbol{H}} \over {\partial\alpha}}
=Na_0^{3/2}e^{T/2}\sqrt{\lambda(\lambda+\alpha)(\lambda+\beta)}
\biggl[-\Bigl({1 \over {\lambda+\alpha}}\Bigr)^2-{1 \over {\lambda(\lambda+\alpha)^2(\lambda+\beta)}}
\Bigl({{G\pi^2} \over 2}\Bigr)\biggr],
\end{eqnarray}

\begin{eqnarray}
\label{DYNAMICS5}
\dot{Y}=G{{\partial\boldsymbol{H}} \over {\partial\beta}}
=Na_0^{3/2}e^{T/2}\sqrt{\lambda(\lambda+\alpha)(\lambda+\beta)}
\biggl[-\Bigl({1 \over {\lambda+\beta}}\Bigr)^2-{1 \over {\lambda(\lambda+\alpha)(\lambda+\beta)^2}}
\Bigl({{G\pi^2} \over 2}\Bigr)\biggr].
\end{eqnarray}

\noindent
and

\begin{eqnarray}
\label{DYNAMICS555}
\dot{T}=G{{\partial\boldsymbol{H}} \over {\partial\lambda}}
=Na_0^{3/2}e^{T/2}\sqrt{\lambda(\lambda+\alpha)(\lambda+\beta)}
\biggl[-\Bigl(\Bigl({1 \over \lambda}\Bigr)^2+({1 \over {\lambda+\alpha}}\Bigr)^2+\Bigl({1 \over {\lambda+\beta}}\Bigr)^2\Bigr)\nonumber\\
-\Bigl({1 \over {\lambda^2(\lambda+\alpha)(\lambda+\beta)}}+{1 \over {\lambda(\lambda+\alpha)^2(\lambda+\beta)}}+{1 \over {\lambda(\lambda+\alpha)(\lambda+\beta)^2}}\Bigr)
\Bigl({{G\pi^2} \over 2}\Bigr)\biggr].
\end{eqnarray}

\noindent
Transferring the exponential to the left hand side and using the identity

\begin{eqnarray}
\label{DYNAMICS51}
e^{-T/2}\dot{T}=-2{d \over {dt}}e^{-T/2},
\end{eqnarray}

\noindent
equation (\ref{DYNAMICS5}) can be written as

\begin{eqnarray}
\label{DYNAMICS51}
{d \over {dt}}e^{-T/2}=
-{1 \over 2}Na_0^{3/2}\sqrt{\lambda(\lambda+\alpha)(\lambda+\beta)}
\biggl[\Bigl(\Bigl({1 \over \lambda}\Bigr)^2+({1 \over {\lambda+\alpha}}\Bigr)^2+\Bigl({1 \over {\lambda+\beta}}\Bigr)^2\Bigr)\nonumber\\
\Bigl({1 \over {\lambda^2(\lambda+\alpha)(\lambda+\beta)}}+{1 \over {\lambda(\lambda+\alpha)^2(\lambda+\beta)}}+{1 \over {\lambda(\lambda+\alpha)(\lambda+\beta)^2}}\Bigr)
\Bigl({{G\pi^2} \over 2}\Bigr)\biggr].
\end{eqnarray}

\noindent
The Hamiltonian constraint comes from the equation of motion for the lapse function $N$, given by

\begin{eqnarray}
\label{DYNAMICS6}
{{\partial\boldsymbol{H}} \over {\partial{N}}}
={1 \over G}a_0^{3/2}e^{T/2}\sqrt{\lambda(\lambda+\alpha)(\lambda+\beta)}\nonumber\\
\biggl[\Bigl(\Bigl({\Lambda \over {a_0^3}}\Bigr)e^{-T}
+\Bigl({1 \over \lambda}+{1 \over {\lambda+\alpha}}+{1 \over {\lambda+\beta}}\Bigr)\Bigr)\nonumber\\
+{1 \over {\lambda(\lambda+\alpha)(\lambda+\beta)}}\Bigl({{G\pi^2} \over 2}\Bigr)\biggr]=0.
\end{eqnarray}

\noindent
Since the pre-factors are assumed classically to be nondegenerate, this implies that the term in square brackets in (\ref{DYNAMICS6}) must vanish, which leads to the condition

\begin{eqnarray}
\label{DYNAMICS7}
{1 \over {\lambda(\lambda+\alpha)(\lambda+\beta)}}
\Bigl({{G\pi^2} \over 2}\Bigr)
=-\biggl[\Bigl({\Lambda \over {a_0^3}}\Bigr)e^{-T}+\Bigl({1 \over \lambda}+{1 \over {\lambda+\alpha}}+{1 \over {\lambda+\beta}}\Bigr)\biggr].
\end{eqnarray}

\noindent
Next we will compute the Hamilton's equations of motion for the gravitational momenta.  These are given by

\begin{eqnarray}
\label{DYNAMICS8}
{1 \over G}\dot{\alpha}=-{{\partial\boldsymbol{H}} \over {\partial{X}}}=0;\nonumber\\
{1 \over G}\dot{\beta}=-{{\partial\boldsymbol{H}} \over {\partial{Y}}}=0;\nonumber\\
{1 \over G}\dot{\lambda}=-{{\partial\boldsymbol{H}} \over {\partial{T}}}
=-{1 \over G}Na_0^{3/2}e^{T/2}\sqrt{\lambda(\lambda+\alpha)(\lambda+\beta)}\nonumber\\
\biggl[-\Bigl({\Lambda \over {a_0^3}}\Bigr)e^{-T}+\Bigl({1 \over \lambda}+{1 \over {\lambda+\alpha}}+{1 \over {\lambda+\beta}}\Bigr)\nonumber\\
+{1 \over {\lambda(\lambda+\alpha)(\lambda+\beta)}}\Bigl({{G\pi^2} \over 2}\Bigr)\biggr].
\end{eqnarray}

\noindent
From (\ref{DYNAMICS8}) one sees that $\alpha$ and $\beta$ are arbitrary complex constants of the motion, independent of time.  Since $\alpha$ and $\beta$ in the quantum theory act as coherent state labels, then the classical dynamics is consistent with the preservation of the coherent states in time.\par
\indent  
Using the Hamiltonian constraint (\ref{DYNAMICS7}), the term in large square brackets in the third equation of (\ref{DYNAMICS8}) is double the first term on account of cancellation of $\hbox{tr}\widetilde{\Psi}^{-1}$.  Hence the equation of motion for $\lambda$ reduces to

\begin{eqnarray}
\label{DYNAMICS9}
\dot{\lambda}=2\Lambda{a}_0^{-3/2}N\sqrt{\lambda(\lambda+\alpha)(\lambda+\beta)}e^{-T/2}.
\end{eqnarray}

\noindent
In constrast to the case for $\alpha$ and $\beta$, it is inappropriate to regard $\lambda$ as a coherent state label, since it is not preserved in time by 
the equations of motion for $\Lambda\neq{0}$.\footnote{This makes sense, since $(\alpha,\beta)$ in the quantum theory play the role of state labels conjugate to $(X,Y)$, which represent the gravitational degrees of freedom.  On the other hand $\lambda$ is conjugate to $T$, which is a time variable on configuration space $\Gamma$.  Note for $\Lambda=0$ that $\lambda$ is independent of time, but for $\Lambda\neq{0}$ it is time-dependent.  This mimics the behavior in the quantum theory.  Also, note that in the instanton representation, all configuration space dependence within the Hamiltonian constraint is confined to $T$.}  The time variation of the $(\lambda,T)$ part of the phase space is determined 
by (\ref{DYNAMICS51}) and (\ref{DYNAMICS9}), which must be solved simultaneously and in conjunction with (\ref{DYNAMICS4}), (\ref{DYNAMICS5}) and (\ref{DYNAMICS555}).

\subsection{Equations of motion for the scalar field}

\noindent
Note that the equations of motion for $X$, $Y$ and $T$ are driven by the scalar field kinetic energy.  But we should also expect a back-reaction from gravity on $\phi$, which we can think of as an inflaton field.  The Hamilton's equations of motion for $\phi$ are given by

\begin{eqnarray}
\label{SCALAR}
\dot{\phi}={{\partial\boldsymbol{H}} \over {\partial\pi}}=Na_0^{3/2}e^{T/2}{1 \over {\sqrt{\lambda(\lambda+\alpha)(\lambda+\beta)}}}\pi;\nonumber\\
\dot{\pi}=-{{\partial\boldsymbol{H}} \over {\partial\phi}}=-Na_0^{-3/2}e^{-T/2}\sqrt{\lambda(\lambda+\alpha)(\lambda+\beta)}V^{\prime},
\end{eqnarray}

\noindent
where $V^{\prime}={{dV} \over {d\phi}}$.  The first line of (\ref{SCALAR}) can be written as

\begin{eqnarray}
\label{SCALAR3}
\pi={{\sqrt{\lambda(\lambda+\alpha)(\lambda+\beta)}} \over {Na_0^{3/2}e^{T/2}}}\dot{\phi},
\end{eqnarray}

\noindent
which implies that

\begin{eqnarray}
\label{SCALAR31}
{{\dot{\pi}} \over {\pi}}=-{{N^2V^{\prime}} \over {\dot{\phi}}}.
\end{eqnarray}

\noindent
We would like to obtain an equation of motion for $\phi$, which we can do by eliminating $\pi$.  Taking the logarithm of the first line of (\ref{SCALAR}), we have
 
\begin{eqnarray}
\label{SCALAR1}
\hbox{ln}\dot{\phi}=\hbox{ln}N+{T \over 2}+\hbox{ln}\pi-{1 \over 2}\bigl(\hbox{ln}\lambda+\hbox{ln}(\lambda+\alpha)+\hbox{ln}(\lambda+\beta)\bigr).
\end{eqnarray}

\noindent
The time derivative of (\ref{SCALAR1}) is given by

\begin{eqnarray}
\label{SCALAR2}
{{\ddot{\phi}} \over {\dot{\phi}}}={{\dot{N}} \over N}+{{\dot{T}} \over {2T}}+{{\dot{\pi}} \over \pi}-{{\dot{\lambda}} \over 2}\Bigl({1 \over \lambda}+{1 \over {\lambda+\alpha}}+{1 \over {\lambda+\beta}}\Bigr).
\end{eqnarray}

\noindent
Substituting (\ref{SCALAR31}) into (\ref{SCALAR2}) and rearranging, we obtain the following equation for $\phi$

\begin{eqnarray}
\label{SCALAR4}
\ddot{\phi}+\Bigl({{\dot{\lambda}} \over 2}\Bigl({1 \over \lambda}+{1 \over {\lambda+\alpha}}+{1 \over {\lambda+\beta}}\Bigr)
-{{\dot{N}} \over N}-{{\dot{T}} \over {2T}}\Bigr)\dot{\phi}+N^2V^{\prime}=0.
\end{eqnarray}

\noindent
Using (\ref{DYNAMICS51}) and (\ref{DYNAMICS9}), equation (\ref{SCALAR4}) can be written in the form

\begin{eqnarray}
\label{SCALAR41}
\ddot{\phi}+3H\dot{\phi}+N^2V^{\prime}=0,
\end{eqnarray}

\noindent
which has the same form as the equation of motion for an inflaton field in Friedman cosmology where $H$ is the Hubble parameter.  But what we have in (\ref{SCALAR41}) is a generalization, which includes the gravitational dynamics predicted by the quantum theory, which is more general than the isotropic case.  Note that $H$ depends on $\lambda(t)$ and $N(t)$ and also is labelled by $\alpha$ and $\beta$.  A possible approach is to perform a Taylor expansion for $\phi$

\begin{eqnarray}
\label{SCALAR5}
\phi(t)=\phi_0+\dot{\phi}_0t+{1 \over 2}\ddot{\phi}_0t^2+{1 \over 6}\dddot{\phi}_0t^3+\dots.
\end{eqnarray}

\noindent
For times small in relation to the timescale of cosmic evolution, only the first few terms of (\ref{SCALAR5}) are important.  The initial values $\phi_0$, $\dot{\phi}_0$, $\lambda_0$ and $N_0$ are freely specifiable, as are the gravitational labels $\alpha$ and $\beta$.  The `new' physics comes in via the second order term of (\ref{SCALAR5}), which can be expressed through the gravitational equations in terms of all values at $t=0$, and likewise for all higher-order terms.  The evolution of $\phi$ also acquires the labels $\alpha$ and $\beta$, and is driven by $N$.

\newpage

\section{Discussion}

\noindent
This paper constitutes a first step in the explicit construction of a Hilbert space for gravity to the Klein--Gordon scalar field for arbitrary self-interaction potential $V(\phi)$.  We have specialized a new procedure using the instanton representation, which generates a gravitational Hilbert space of normalizable states, to minisuperspace.  The next natural step will be the generalization of this procedure to additional matter fields, first in minisuperspace and then to progress to the full theory.  The ultimate goal is to apply this procedure to the Standard Model.

\section{Appendix: Functional Green's functions}

\noindent
To explicitly compute $\Lambda^n_{\alpha,\beta,p}$ one makes use of the identity

\begin{eqnarray}
\label{PAIR}
f(\partial/\partial\rho)\delta(\rho)=F(\rho),
\end{eqnarray}

\noindent
where $f$ and $F$ are a Laplace transform pair, related by

\begin{eqnarray}
\label{PAIR1}
f(\rho)=\int^{\infty}_0e^{-\rho{z}}F(z)dz.
\end{eqnarray}

\noindent
Applying this to the operator $\hat{O}$, we have

\begin{eqnarray}
\label{PAIR2}
\Bigl({{\partial^2} \over {\partial\rho^2}}+O(\lambda_{\alpha,\beta,p}-k\mu)\Bigr)^{-1}\delta(\rho)
={{i\hbox{sin}\sqrt{O(\lambda_{\alpha,\beta,p}-k\mu)}\rho} \over {2\sqrt{O^k_{\alpha,\beta,p}}}}.
\end{eqnarray}

\noindent
The index $k$ labels the particular order in the expansion, which due to $e^{-T}$ is lowered by one unit for each iteration.  The $(\alpha,\beta)$ labels come from the gravitational degrees of freedom which on the constrained space start out orthogonal to $\lambda$.  
The action on the cosmological term involves the convolution

\begin{eqnarray}
\label{PAIR3}
\Bigl({{\partial^2} \over {\partial\rho^2}}+O(\lambda_{\alpha,\beta,p}-k\mu)\Bigr)^{-1}\Lambda(\rho)
=\int{d\rho^{\prime}}{{i\hbox{sin}\sqrt{O(\lambda_{\alpha,\beta,p}-k\mu)}(\rho-\rho^{\prime})} \over {2\sqrt{(\lambda_{\alpha,\beta,p}-k\mu)}}}\Lambda(\rho^{\prime}).
\end{eqnarray}

\noindent
The result to $n^{th}$ order is given by

\begin{eqnarray}
\label{PAIR4}
\Lambda^n_{\alpha,\beta,p}
=(i/2)^n\prod_{k=1}^n{{Q(\lambda_{\alpha,\beta,p}-k\mu)} \over {\sqrt{O(\lambda_{\alpha,\beta,p}-k\mu)}}}\nonumber\\
\int{d\rho_n}\int{d}\rho_{n-1}\dots\int{d\rho_2}\int{d\rho_1}\prod_{l=1}^n\Lambda(\rho_l)\hbox{sin}\bigl[\sqrt{O(\lambda_{\alpha,\beta,p}-l\mu)}(\rho_{l+1}-\rho_l)\bigr].
\end{eqnarray}

\noindent
As a matter of practical calculation for specific potentials, one could also use the identity

\begin{eqnarray}
\label{PAIR5}
\Bigl({{\partial^2} \over {\partial\rho^2}}-K\Bigr)^{-1}=\Bigl({\partial \over {\partial\rho}}-\sqrt{K}\Bigr)^{-1}\Bigl({\partial \over {\partial\rho}}+\sqrt{K}\Bigr)^{-1}
\end{eqnarray}

\noindent
This has the interpretation of `shifted' integration.  The action on the cosmological term is given by

\begin{eqnarray}
\label{PARI6}
e^{\sqrt{K}\rho}\int^{\rho}d\rho^{\prime}e^{-2\sqrt{K}\rho}\int^{\rho^{\prime}}d\rho^{\prime\prime}e^{\sqrt{K}\rho^{\prime\prime}}\Lambda(\rho^{\prime\prime}),
\end{eqnarray}

\noindent
which is more straightforward to use for polynomial potentials, exponential and numerically constant in $\rho$.

\end{document}